\documentclass[twocolumn,prb,aps,showpacs]{revtex4}
\usepackage{graphicx}
\usepackage{color}
\usepackage{dcolumn}
\usepackage{amsmath}
\newlength{\figwidth}
\newlength{\figwidthb}
\newcommand{\vc}[1]{\mbox{\boldmath $#1$}} 
\newcommand{\ei}{E_{\mathrm i}} 
\newcommand{\ef}{E_{\mathrm f}} 
\newcommand{\etr}{\varepsilon} 
\newcommand{\fb}{FeBO$_{3}$} 
\newcommand{\stup}{1$\underline{s}$-4$p$} 
\newcommand{\stud}{1$\underline{s}$-3$d$} 

\newcommand{\gapmh}{U_{_{\mathrm {G}}}} 
\newcommand{\gapct}{\Delta_{_{\mathrm {G}}}} 
\begin{document}

\title{Charge-transfer and Mott-Hubbard Excitations in \fb: Fe K edge Resonant Inelastic
  X-ray Scattering Study}

\author{Jungho Kim}\email{jhkim@aps.anl.gov} \affiliation{Advanced Photon
  Source, Argonne National Laboratory, Argonne, Illinois 60439, USA}
\author{Yuri Shvyd'ko}\email{shvydko@aps.anl.gov} \affiliation{Advanced Photon Source,
  Argonne National Laboratory, Argonne, Illinois 60439, USA}

\date{\today}

\begin{abstract}
{\em Momentum-resolved} resonant inelastic x-ray scattering (RIXS)
spectroscopy has been carried out at the Fe K edge, successfully for
the first time. The RIXS spectra of a \fb\ single crystal reveal a
wealth of information on $\simeq 1-10$~eV electronic excitations.
The IXS signal resonates both when the incident photon energy
approaches the pre-edge (1$\underline{s}$-3$d$) and the main-edge
(1$\underline{s}$-4$p$) of the Fe K edge absorption spectrum. The
RIXS spectra measured at the pre-edge and the main-edge show
quantitatively different dependences on the incident photon energy,
momentum transfer, polarization, and temperature. Electronic
excitations observed in the pre-edge and main-edge RIXS spectra are
interpreted as Mott-Hubbard (MH) and charge-transfer (CT)
excitations, respectively. The charge-transfer gap $\gapct
=3.81\pm0.04$~eV and the Mott-Hubbard gap $\gapmh = 3.96\pm0.04$~eV,
are determined, model independently from the experimental data. The
CT and MH excitations are assigned using molecular orbitals (MO) in
the cluster model and multiplet calculation in the many-electron
multiband model, respectively.
\end{abstract}

\pacs{78.70.En, 71.20.Be, 78.20.-e, 75.50.Ee}

\maketitle

\section{Introduction}

Iron borate, \fb\, is a classical material with strong electron
correlations experiencing interesting magnetic, optical, and
magneto-optical properties. A review of the early studies on crystal
growth, structure, physical properties (magnetic, elastic,
magnetoelastic, optical, and magnetooptical) and applications of
\fb\ can be found in Ref.~\onlinecite{DJN84}. It is among only a few
known materials magnetically ordered at room temperature and
transparent in the visible spectrum, \cite{Kurtzig69,Wolfe70} making
\fb\ attractive in applications for broadband visible
magneto-optical devices. A number of electronic and magnetic phase
transitions induced by high pressure have been observed recently:
collapse of the magnetic moment of Fe$^{3+}$ ions,
\cite{TGS01,Sarkisyan02} insulator-semiconductor transition,
\cite{Troyan03,GTO04} increase of the Ne\'el temperature.
\cite{GTL05} The driving mechanism for these high-pressure phenomena
is attributed to the electronic transition in Fe$^{3+}$ ions from
the high-spin 3$d^{5} (S=5/2,^{6}A_{1g})$ to the low-spin
$(S=1/2,^{2}T_{2g})$ state (spin crossover).~\cite{GTL05,Lyubutin09}

Despite a great number of experimental and theoretical studies that
have been performed, the electronic structure of \fb\ is still not
completely understood. Ovchinnikov and Zabluda~\cite{OZ04} developed
an empirical many-electron model that took all $d$ orbitals into
account and strong electron correlations involving $d$ electrons of
Fe atoms. The magnitude of the main parameters of the band
structure, such as the charge-transfer (CT) gap $\gapct$,
Mott-Hubbard (MH) gap $\gapmh$, crystal-field splitting $10Dq$,
etc., are derived from optical absorption spectra and x-ray
photoemission spectroscopy. \cite{OZ04,GTO04} The same scheme has
been applied to interpret high pressure induced magnetic collapse
and insulator-metal transition. \cite{Lyubutin09} The
first-principles local-density approximation (LDA) energy band
calculations predicted an antiferromagnetic (AFM) metal instead of
an AFM insulator. \cite{Postnikov94} Shang $et$~$al.$ \cite{Shang07}
have performed first-principles band-structure calculations using
the density functional theory within the generalized gradient
approximation (GGA) and the GGA+U approach. The electronic structure
was predicted to be high-spin, antiferromagnetic and insulating, in
agreement with experiments. However, in order to predict the correct
value of the bandgap in the first-principle calculation, the Coulomb
repulsion $U=7$~eV has to be assumed, much higher than $U=2.97$~eV
obtained in the empirical many-electron model of
Ref.~\onlinecite{OZ04}.

An insulating ground state in a transition metal (TM) compound is
realized by an on-site $U$, giving rise to the split of the
conduction band into lower and upper Hubbard bands (LHB and UHB,
respectively). If $U$ is small, the insulating charge gap arises
from the MH excitation between LHB and UHB. On the other hand, when
$U$ is quite large, the insulating charge gap arises from the CT
excitation between the highest occupied ligand ($O$~2$p$) and UHB.
In the Zaanen-Sawatzky-Allen (ZSA) scheme,~\cite{ZSA85} the TM
compound is classified as a MH type in the first case and a CT type
in the second case. The general trend is that early TM compounds
belong to the MH type and late TM compounds, such as cuprates,
belong to the CT type. For Fe compounds with half-filled ($d^5$),
such as \fb, however, the classification is not as clear; $\gapmh$
from the lowest MH excitation is comparable with $\gapct$ from the
lowest CT excitation.

%
%
\begin{figure*}[t]
\setlength{\unitlength}{\textwidth}
\begin{picture}(1,0.45)(0,0)
\put(0.01,0.01){\includegraphics[width=0.95\textwidth,angle=0]{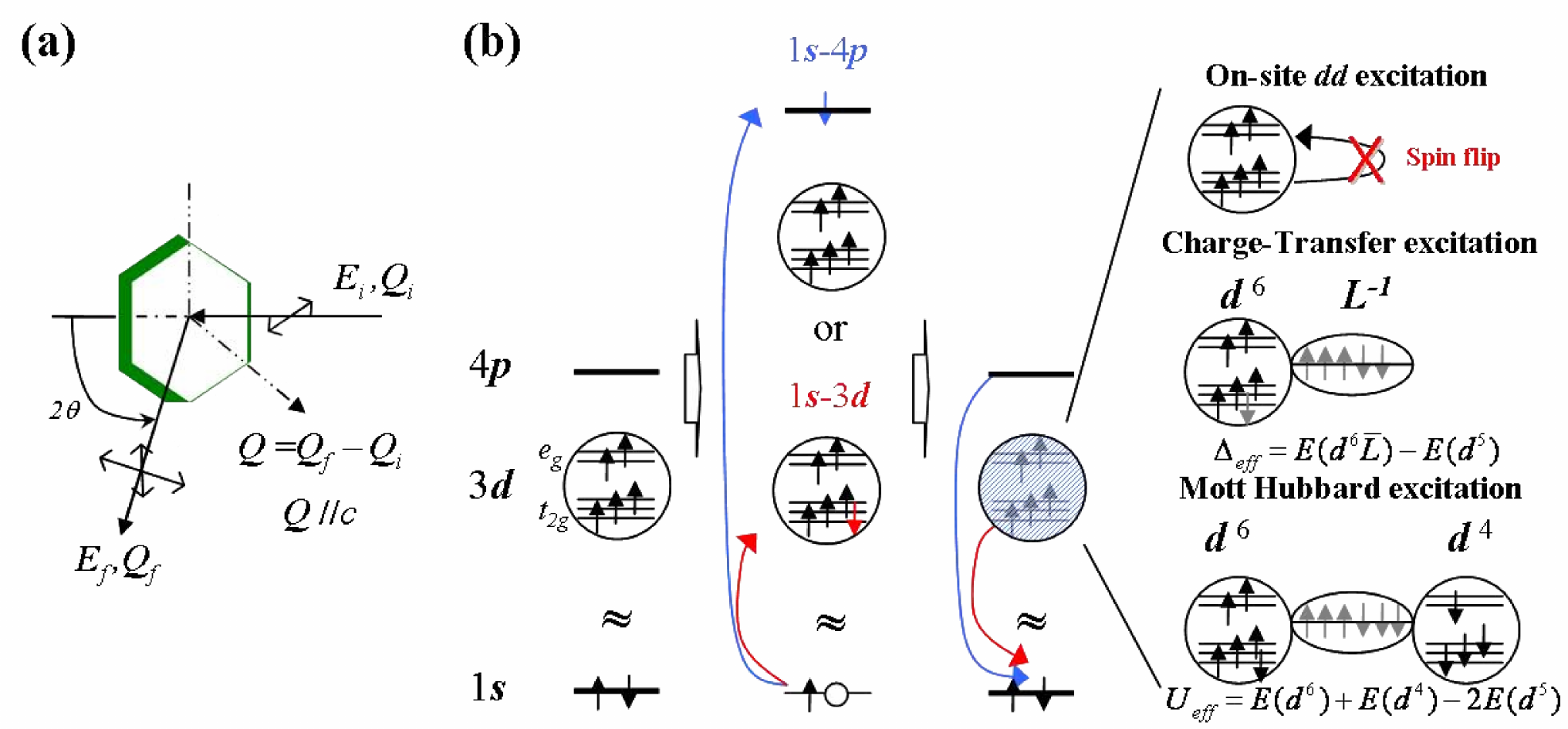}}
\end{picture}
\caption{{(a) Schematic of the RIXS experiment in the horizontal
scattering geometry, where the incident photon polarization is
parallel to the scattering plane ($\pi$-polarization). The \fb\
sample is mounted in a way that the momentum transfer
$\vc{Q}=\vc{Q}_{\mathrm f}-\vc{Q}_{\mathrm i}$ is along the crystal
$\mathbf{c}$ axis. The scattered photon energy is measured by a
spectrometer consisting of a Ge$(620)$ spherical diced analyzer and
a position-sensitive micro strip detector. (b) Schematic of a Fe K
edge RIXS process in \fb. In the initial state, five 3$d$ electrons
fill $t_{2g}$ and $e_{g}$ crystal-field levels in the high spin
configuration. Two distinct absorption transitions can be used for
RIXS measurement: the pre-edge (\stud) and the main-edge (\stup) of
the Fe K edge absorption spectrum. Various valence electronic
excitations are left after the decay of the excited electron back
into the 1$s$ core level: on-site $dd$, charge-transfer, and
Mott-Hubbard excitations. On-site $dd$ excitations are forbidden
because they require spin-flip.}}\label{fig:intro}
\end{figure*}

Probing CT and MH excitations is important to resolve the existing
ambiguities in determination of the electronic structure and
parameters in \fb. Conventional spectroscopy, such as optical
spectroscopy, is of limited utility; it has difficulty in probing
dipole-forbidden charge excitation, such as MH excitation, and
characterization of probed charge excitation is not straightforward
because it cannot probe CT and MH excitations selectively. A higher
order spectroscopy that uses the excited electronic state of
interest as an intermediate state can provide a way to circumvent
the limitation.

Momentum-resolved resonant inelastic x-ray scattering (RIXS) at the
K edge plays an increasingly important role as a photon-in
photon-out higher order spectroscopic tool for investigation of
localized and propagating charge excitations in TM compounds
providing bulk-sensitive, element-specific information. A variety of
systems, including
cuprates,~\cite{Abbamonte99,Hasan00,Yjkim02,Yjkim04,Ishii05,Suga05,Lu05}
manganites,~\cite{Inami03,Grenier05} and
nickelates,~\cite{Collart06, WKI09} have been studied by RIXS.
Charge excitations probed by RIXS are known to strongly depend on
the RIXS intermediate excited state. The CT excitations are usually
observed in cuprates by K edge
RIXS.~\cite{Hasan00,Yjkim02,Yjkim04,Ishii05,Lu05,yjkim07,JKim09,Collart06}
On the other hand, on-site $dd$ excitations of cuprates can be
probed by RIXS at Cu $M_{2,3}$- \cite{Kuiper98} and $L_3$-edge,
\cite{Ghiringhelli04} which involves direct transitions to and from
3$d$ states.

It is well known that there exist two distinct transitions in the Fe
K edge: the pre-edge \stud\ and the main-edge \stup\, where the
underline denotes a hole. While the latter is dipole allowed, the
former \stud\ transition is quadrupole allowed or particularly
dipole allowed when static or dynamic local distortions exist. The
\stud\ absorption would show up as a weak peak, just below the
strong \stup\ absorption peak. This pre-edge (\stud) can be used for
the resonant enhancement of an inelastic x-ray scattering (IXS)
signal. The pre-edge RIXS enables direct access to the $3d$ valence
system, similar to the M edge (3$\underline{p}$-3$d$) or the L edge
(2$\underline{p}$-3$d$) RIXS \cite{Schuelke07} but without a
complexity imposed by the spin-orbit interaction.

%
%
\begin{figure*}[t]
\setlength{\unitlength}{\textwidth}
\begin{picture}(1,0.51)(0,0)
\put(0.01,-0.02){\includegraphics[width=1\textwidth]{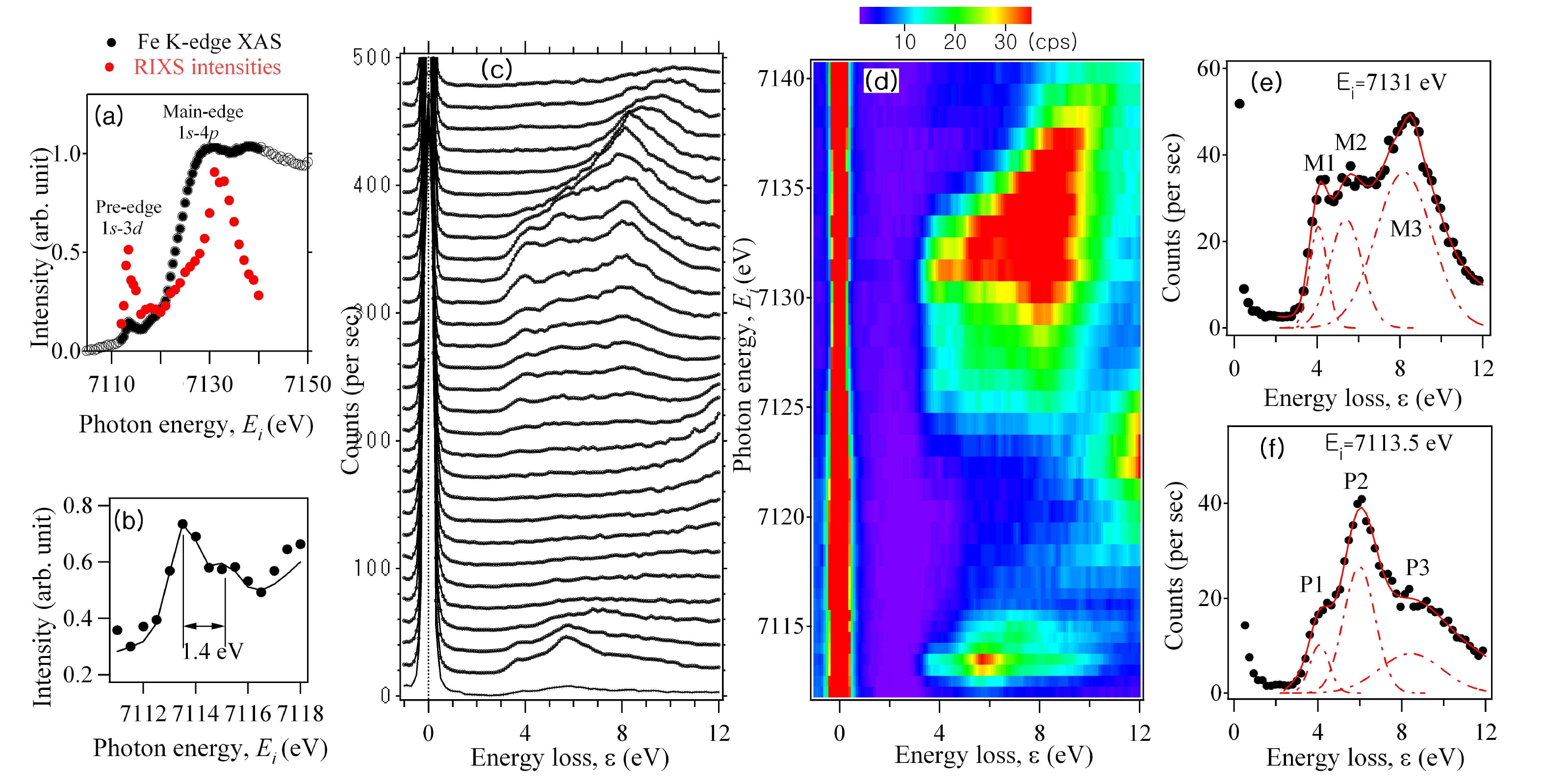}}
\end{picture}
\caption{{(a) Fe K edge x-ray absorption spectrum (XAS) of \fb\
measured in PFY mode (black open dots) and integrated intensities of
the RIXS spectra of (c) (red filled dots). (b) XAS same as in (a)
but shown on the expanded scale of $\ei$ around the pre-edge. (c)
RIXS spectra measured for different incident photon energies
$7112~{\mathrm{eV}}\le\ei\le 7140$~eV. (d) 2D plot of the RIXS
spectra of (c). (e) and (f) show representative RIXS spectra of the
main-edge RIXS ($E_{i}$=7131~eV, Q=(0~0~9)) and the pre-edge RIXS
($E_{i}$=7113.5~eV, Q=(0~0~9)), respectively. Gaussian functions are
used to fit inelastic peaks, while the Voigot function is used to
fit the elastic line (see the text).}}\label{fig:energydep}
\end{figure*}

Figure~\ref{fig:intro}(b) shows a schematic diagram of the Fe K edge
RIXS process in \fb. In the initial state, five 3$d$ electrons fill
$t_{2g}$ and $e_{g}$ crystal-field levels in the high-spin
configuration. As a first step, the 1$s$ core electron is excited to
3$d$ (pre-edge) or 4$p$ (main-edge), the excited electron decays
back into the 1$s$ core level creating various charge excitations
with an energy $\etr$ equal to the energy loss of the incident
photon $\etr=\ei-\ef$. Measuring RIXS spectra $S(\etr,\vc{Q},\ei)$
for a given momentum transfer $\vc{Q}=\vc{Q}_{\mathrm
f}-\vc{Q}_{\mathrm i}$ provides valuable information on charge
excitations in solids. The availability of two absorption
transitions at the K edge provide an opportunity to probe charge
excitations selectively. One can assume that the on-site $dd$ and MH
excitations could be better observed in the pre-edge RIXS than the
main-edge RIXS, because an intermediate state of the pre-edge RIXS
directly involves 3$d$ states. However, until now this has not been
demonstrated.

In the present study, we explore this possibility by measuring RIXS
spectra using both intermediate states. We will refer to the
pre-edge RIXS when the incident photon energy is tuned to the energy
of the \stud\ ($\ei \approx 7113.5$~eV) transition. Similarly, we
will refer to the main-edge RIXS if the incident photon energy is
tuned to the energy of the \stup\ ($\ei \approx 7131$~eV)
transition. In the following sections, we will present the pre-edge
and main-edge RIXS spectra in \fb\ measured as a function of the
excitation energy, momentum transfer, polarization, and temperature.
We show that unlike previously studied cases, the K edge RIXS can
probe both CT and MH excitations at the same time. We interpret the
pre-edge and the main-edge RIXS excitations as MH and CT
excitations, respectively, and derive important quantities including
the charge-transfer gap $\gapct$ and the Mott-Hubbard gap $\gapmh$
using molecular orbitals (MO) in the cluster model and multiplet
calculation in the many-electron multiband model. This study
demonstrates a special quality of the K edge RIXS to be a
simultaneous probe of CT and MH excitations in TM compounds.

\section{Experiment and Sample}

The RIXS measurements were performed using the MERIX spectrometer at
the XOR-IXS 30-ID beamline of the Advanced Photon Source (APS). The
sample is mounted in the Displex closed-cycle cryostat NE-202N with
a temperature range of 6~K to 450~K. The measurements were carried
out at room temperature and at temperatures close to the Ne\'el
temperature ($T_{\mathrm N}$=348~K). X-rays impinging upon the
sample were monochromatized to a bandwith of $75$~meV, using a
four-bounce $(+--+)$ monochromator with asymmetrically cut Si(400)
crystals.~\cite{ToellnerMERIX} The beam size on the crystal was
reduced to 45(H)$\times$20(V)~$\mu$m$^2$ by focusing in the
Kirkpatrick-Baez configuration. The photon flux on the sample was
$1.1 \times 10^{12}$~ph/s. The total energy resolution of the MERIX
spectrometer at the Fe K edge is $180$~meV. This is achieved using a
Ge$(620)$ spherical diced analyzer, and a position-sensitive
microstrip detector placed on a Rowland circle with a $1$~m radius.
The silicon microstrip detector with $125$~$\mu$m pitch is applied
for the purpose of reducing the geometrical broadening of the
spectral resolution function.~\cite{HVA05} Maximum RIXS count rates
were in the range of 40-50~Hz. Horizontal scattering geometry, where
the incident photon polarization vector component is parallel to the
scattering plane ($\pi$-polarization), was used for all RIXS
measurements, with the crystal $\mathbf{c}$-axis in the scattering
plane as shown in Fig.~\ref{fig:intro}(a).

Iron borate, \fb, single crystal with low-dislocation density, grown
by spontaneous crystallization from flux,~\cite{KKN85} was used in
the current experiment. The crystal has a form of a platelet
$6\times 7\times 0.15$~mm$^3$, with $\mathbf{c}$-axis perpendicular
to the platelet surface. Iron borate has a rhombohedral calcite
structure that belongs to the space group $R\overline{3}c(D^6_{3d})$
with two formula units per unit cell. The lattice constants are
$\mathbf{a}$=$\mathbf{b}$=4.626(1)~${\AA}$ and
$\mathbf{c}$=14.493(6)~${\AA}$. The Fe$^{3+}$ ions are centered in a
slightly distorted $O^{2+}_{6}$ octahedra. The octahedral $O_{h}$
crystal field splits the energy of the $3d$ orbitals into three-fold
degenerate $t_{2g}$ and two-fold degenerate $e_g$ states. As
depicted in Fig.~\ref{fig:intro}(b), five $3d$ electrons fill these
crystal-field levels in the high-spin
configuration.~\cite{Sarkisyan02} \fb\ is a large-gap
antiferromagnetic insulator with the N\'eel temperature $T_{\mathrm
N}$=348~K below which the two sublattice magnetic moments along the
$\mathbf{c}$-axis order antiferromagnetically. All of the Fe$^{3+}$
spins are in the plane perpendicular to the $\mathbf{c}$-axis.
Nearest-neighbor (NN) exchange interaction of the Fe$^{3+}$ spins is
antiferromagnetic. A slight canting of two sublattice magnetic
moments gives rise to a weak ferromagnetic moment.

\section{Data and analysis}

\subsection{Incident photon energy ($\ei$) dependence}

Figure~\ref{fig:energydep}(a) presents Fe K edge x-ray absorption
spectra (XAS) of \fb , measured in the partial fluorescence yield
(PFY) mode, by detecting Fe~K$_{\alpha}$ fluorescence. A strong
absorption main-edge (\stup) around 7130~eV and a weak absorption
pre-edge (\stud) around 7114~eV are observed.
Figure~\ref{fig:energydep}(b) shows the pre-edge XAS on the expanded
energy scale. A fine structure with two peaks can be resolved. The
peak separation is estimated to be 1.4~eV. In the strong crystal
field limit, two peaks are assigned to 1$\underline{s}$-3$d(t_{2g})$
and 1$\underline{s}$-3$d(e_{g})$ transitions. The peak separation
corresponds to the crystal-field splitting,
10$Dq$=1.4~eV.~\cite{tami97}

The RIXS spectra have been measured at $\vc{Q}$=(0~0~9) with $\ei$
changing from the pre-edge to the main-edge absorption energies
indicated by filled black circles in Fig.~\ref{fig:energydep}(a).
Figure~\ref{fig:energydep}(c) shows RIXS spectra measured as a
function of the energy loss ($\etr$) with 1~eV increment in $\ei$.
Every spectrum shows a strong elastic signal at $\etr$=0. The RIXS
signal resonates around $\ei$=7131~eV and $\ei$=7113.5~eV. The 12~eV
feature in the RIXS spectra observed for $\ei$ between main-edge and
pre-edge RIXS is due to the $K\beta_5$ emission. Intensities of the
RIXS spectra integrated in the range $2~{\mathrm{eV}}\le\etr\le
12$~eV are plotted in Fig.~\ref{fig:energydep}(a). The two
resonances are clearly seen.

Figure~\ref{fig:energydep}(d) shows a 2D plot of RIXS intensities in
($\ei,\etr$) space. That there is a difference in resonance
dependence for the main-edge and the pre-edge RIXS is easily seen.
The 2D color plot in the main-edge RIXS region shows a typical $\ei$
dependence observed in a number of previously reported main-edge
RIXS studies on cuprates:~\cite{Abbamonte99,Lu06,yjkim07} the
intensity of the 4~eV and 6~eV features drops rapidly as $\ei$
increases from 7131~eV. Such resonance behavior is often explained
by the third-order perturbation theory, where the RIXS cross section
can be factorized into a resonant prefactor that depends on the
incident and scattered photon energies and dynamic structure
factor.~\cite{Platzman98,Abbamonte99,Lu06,yjkim07,JKim09} In
contrast, the shape of the pre-edge RIXS spectrum is stable with
respect to change in $\ei$. A similar $\ei$ dependence is observed
in the L edge (2$\underline{p}$-3$d$) RIXS~\cite{Ghiringhelli05} and
the Ni pre-edge (1$\underline{s}$-3$d$) RIXS in NiO, in which the
on-site $dd$ excitations are probed. This observation suggests that
the features in the pre-edge RIXS spectrum undergo the same RIXS
process as with $dd$ excitations, which occur in the direct
transition to and from the 3$d$ states and are well described by the
second-order perturbation theory (Kramers-Heisenberg equation).

The RIXS spectra show no low-energy feature with $\etr<$3~eV. This
is the region where on-site $dd$ excitation should take place.
On-site $dd$ excitations in \fb\ are forbidden when five $3d$
electrons of \fb\ are in the high-spin configuration as shown in
Fig.~\ref{fig:intro}(b). This observation proves that RIXS does not
probe charge excitation accompanied by spin-flip.

Figures~\ref{fig:energydep}(e) and (f) show representative RIXS
spectra corresponding to the main-edge ($E_{i}$=7131~eV) and the
pre-edge ($E_{i}$=7113.5~eV) RIXS, respectively. In order to obtain
quantitative information on the peak positions and their
intensities, Gaussian functions are used to fit inelastic peaks
while the Voigot function is used to fit the elastic peak. Note that
the influence of elastic tail intensity on the inelastic intensity
is marginal because the inelastic peaks are pretty far away from the
elastic peak. For the main-edge RIXS spectra, three peaks could be
identified at around 4.09~eV (M1), 5.29~eV (M2), and 8.14~eV (M3).
The pre-edge RIXS spectra show different high-energy features: a
small shoulder peak at around 3.96~eV (P1), a  strong main peak at
around 5.85~eV (P2) peak, and a broad peak at around 8.71~eV (P3).

\subsection{Momentum transfer dependence}

%
%
\begin{figure*}[t]
\setlength{\unitlength}{\textwidth}
\begin{picture}(1,0.49)(0,0)
\put(0.01,-0.09){\includegraphics[width=01\textwidth]{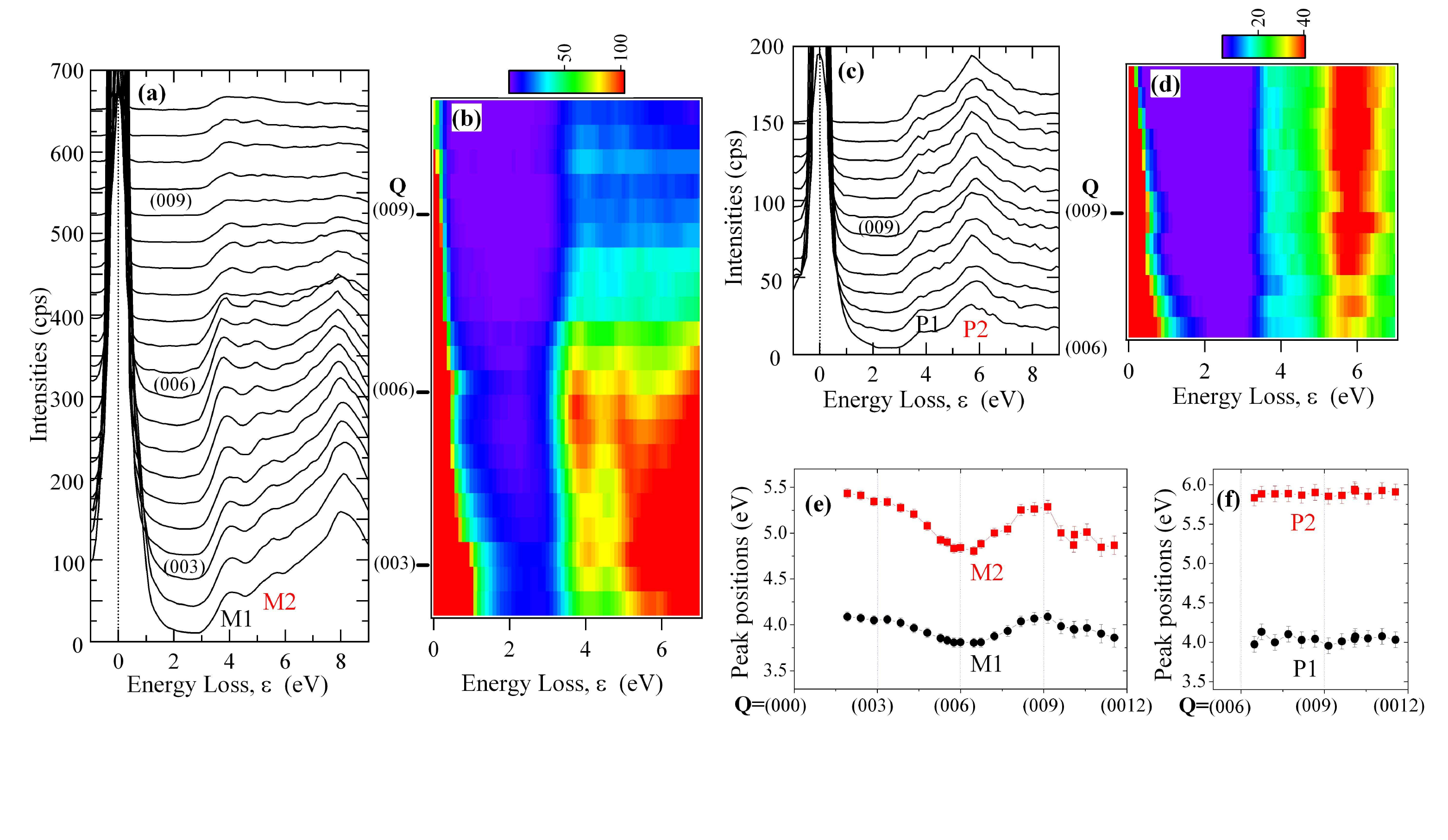}}
\end{picture}
\caption{RIXS spectra measured at the (a) main-edge, and (c)
pre-edge as function of $\vc{Q}=(0~0~L)$, $2<L<11.5$. 2D plots of
the corresponding RIXS data are shown in (b) and (d), respectively.
(e) $\vc{Q}$ dependence of the M1 and M2 peaks derived from RIXS
sepctra in (a). (f) $\vc{Q}$ dependence of the P1 and P2 peaks from
RIXS spectra in (c).}\label{fig:momentum}
\end{figure*}

Momentum transfer ($\vc{Q}$) dependence of the RIXS spectrum
provides valuable information on the dispersion of the occupied and
unoccupied bands.~\cite{Tsutsui03} In this section, we present
$\vc{Q}$ dependences of the observed RIXS spectra measured along the
$\mathbf{c}$-axis: $\vc{Q}=(0~0~L)$.

Figures~\ref{fig:momentum}(a) and (b) show the main-edge RIXS
spectra measured for different $\vc{Q}$. The scattering angle
2$\theta$ of the bottom and the top spectra in
Fig.~\ref{fig:momentum}(a) are about 14$\rm^o$ ($L\cong2$) and
92$\rm^o$($L\cong11.5$), respectively. Noticeable changes with
$\vc{Q}$ are observed in peak positions and intensities. The M1 and
M2 peaks disperse shifting to lower energy as $\vc{Q}$ approaches
(0~0~6) and moving back to higher energy as $\vc{Q}$ increases
further. The RIXS intensity decreases for $L>6$ as shown in
Fig.~\ref{fig:momentum}(b). Let us discuss first the $\vc{Q}$
dependence of the peak positions. The peak intensity changes will be
discussed later.

The positions of M1 and M2 peaks are shown as a function of $\vc{Q}$
in Fig.~\ref{fig:momentum}(e). Peak M1 is located at around 4.09~eV
for $\vc{Q}$=(0~0~1.9). As $\vc{Q}$ approaches (0~0~6), the peak
energy changes to $3.81\pm0.04$~eV. With a further increase of
$\vc{Q}$ to (0~0~9), M1 moves back to the higher energy of
$4.09\pm0.07$~eV. Although it is not evident from the raw data of
Fig.~\ref{fig:momentum}(a) and (b), due to a small RIXS intensity at
$L>9$, the results from the numerical data evaluation shown in
Fig.~\ref{fig:momentum}(e) demonstrate that M1 again disperses to
the lower energy of $3.86\pm0.10$~eV as $\vc{Q}$ approaches
(0~0~12). The dispersion width of the M1 peak is about 0.26~eV. The
M2 peak shows the same dispersion behavior but with a larger
dispersion width of 0.44~eV: $4.84\pm0.04$~eV at $\vc{Q}$=(0~0~6),
$5.29\pm0.07$~eV at $\vc{Q}$=(0~0~9), and $4.87\pm0.10$~eV at
$\vc{Q}$=(0~0~12). Note that $\vc{Q}$=(0~0~6) corresponds to the
first allowed Bragg reflection along this highly symmetric direction
and (0~0~12) is the second one. Therefore, these two low-lying
excitations have the normal dispersion relation showing direct
energy gaps. The mean energy difference between these two low-lying
peaks is $1.13\pm0.13$~eV.

Figures~\ref{fig:momentum}(c) and (d) show the pre-edge RIXS spectra
measured for different $\vc{Q}=(0~0~L)$. The 2$\theta$ values of the
bottom and the top spectra in Fig.~\ref{fig:momentum}(c) are about
48$\rm^o$ ($L\cong6.5$) and 92$\rm^o$ ($L\cong11.5$), respectively.
Contrary to the M1 and M2 peaks in the main-edge RIXS spectra, the
P1 and P2 peaks do not show any noticeable peak intensity and
position changes. The positions of P1 and P2 are shown in
Fig.~\ref{fig:momentum}(f). As mentioned above, the peak positions
are stable. The mean energy is $3.96\pm0.05$~eV and $5.85\pm0.03$~eV
for M1 and M2, respectively. The mean energy difference between
these two low-lying peaks is $1.85\pm0.04$~eV.

The dispersive nature of charge excitation has been reported for a
CT excitation of cuprates in a number of RIXS
studies.~\cite{Hasan00,Yjkim02,Lu05,Collart06,Ishii07,Ellis08} The
CT excitation in cuprates is defined as the excitation from the
Zhang-Rice band (O 2$p$) to the upper Hubbard band (UHB). For
example, the 2~eV peak of La$_2$CuO$_4$ which is assigned as the CT
excitation, shows a sizable dispersion of
0.1$\sim$0.5~eV.~\cite{Lu05,Collart06,Ellis08} On the other hand, it
has been reported that Mott-Hubbard (MH) excitations in manganites
do not show any obvious dispersive
behavior.~\cite{Inami03,Ishii04,Grenier05} The MH excitation in
manganites is defined as the excitation from the lower Hubbard band
(LHB) to the UHB. The dispersive behavior of the CT excitations and
non dispersive behavior of the MH excitation can be attributed to
the fact that, in general, the delocalized ligand 2$p$ band has a
larger bandwidth than the transition metal $3d$ band.

Using this analogy, the charge excitations probed by the main-edge
and pre-edge RIXS can be attributed to CT and MH excitations,
respectively. In these assignments, the dispersion of the main-edge
RIXS excitation is understood as resulting from the dispersive
occupied oxygen 2$p$ band and the dispersive UHB, while no
dispersion of the pre-edge RIXS excitation is attributed to the non
dispersive LHB.

\subsection{Photon polarization dependence}

The intensity of the RIXS features vary with $\vc{Q}$ or equivalent
scattering angle. This can be attributed to the photon polarization
dependence of the RIXS cross section. In our experiment, the
incident photon polarization component ($\vec{\epsilon}_i$) is
parallel to the scattering plane ($\pi$-polarization). The
scattering plane is defined by wave vectors of the incident and the
scattered photons. In this section, we present the photon
polarization dependence of the RIXS spectra in \fb.

2D plots in Fig.~\ref{fig:momentum}(b) and (d) show how the RIXS
intensities change with $\vc{Q}$. First, we note that elastic line
intensities in both cases reduce substantially as $\vc{Q}$
increases, especially, when $2\theta$ approaches 90$\rm^o$
($L\simeq$12). This is related to the fact that the polarization
dependence of the elastic (Thomson) scattering cross section is
proportional to the scalar product
$\vec{\epsilon}_i\cdot\vec{\epsilon}_s$ of $\vec{\epsilon}_i$ and
scattered photon polarization ($\vec{\epsilon}_s$).

The main-edge and the pre-edge RIXS spectra show quite different
inelastic intensity changes with $\vc{Q}$. In the case of the
main-edge RIXS, the intensity of the charge excitations is more or
less stable near $\vc{Q}$=(0~0~6) but rapidly disappears for a
higher $\vc{Q}$, as shown in Fig.~\ref{fig:momentum}(b). On the
other hand, the intensities of the charge excitations in the
pre-edge RIXS do not show any dramatic $\vc{Q}$ dependence.

%
%

\begin{figure}[t]
\setlength{\unitlength}{\textwidth}
\begin{picture}(1,0.34)(0,0)
\put(-0.045,-0.049){\includegraphics[width=0.58\textwidth,angle=0]{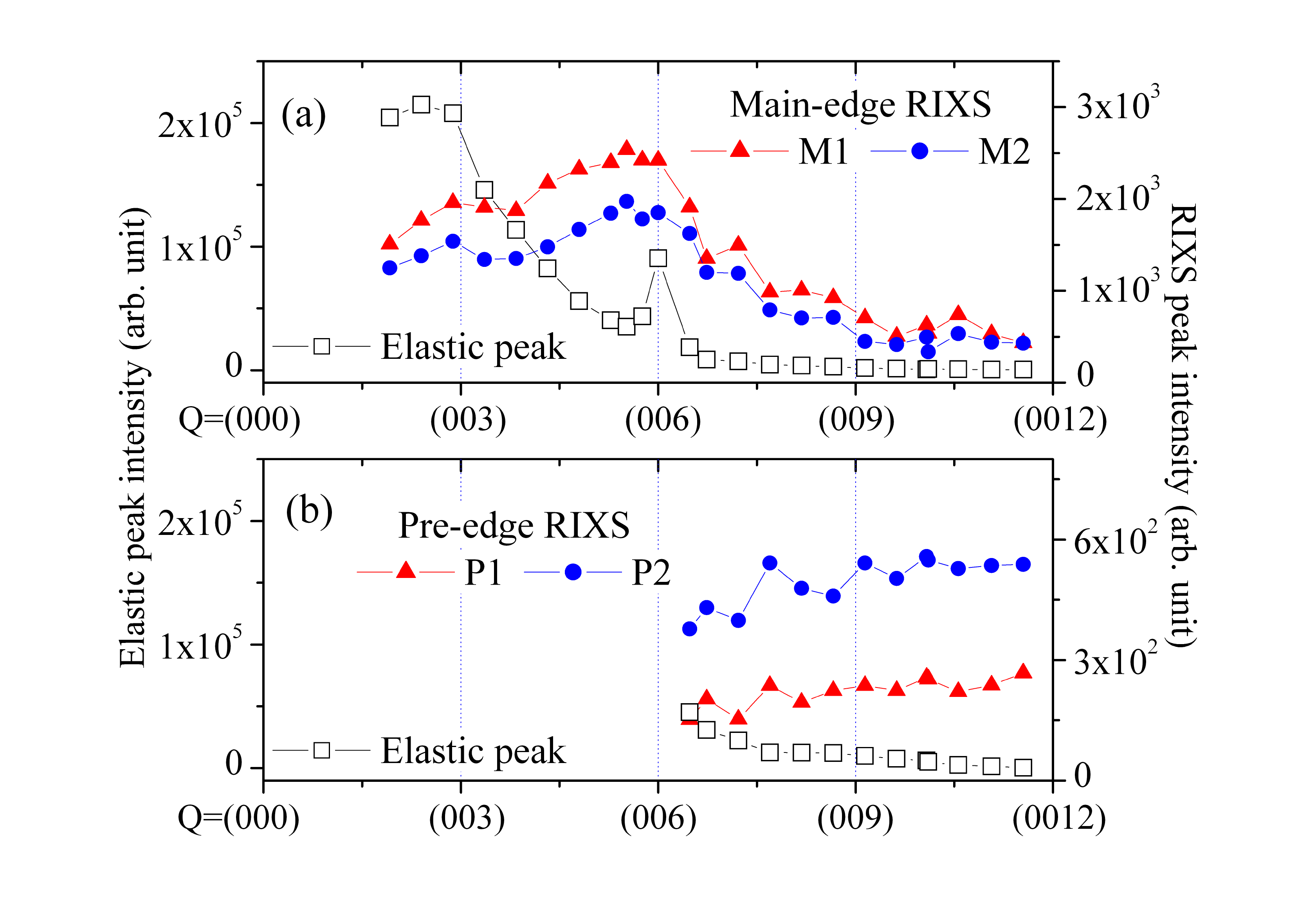}}
\end{picture}
\caption{Intensity of charge excitations measured as a function of
momentum transfer $\vc{Q}$ in (a) main-edge and (b) pre-edge RIXS.
Open square symbols denote the elastic peak intensities, while
filled symbols denote the inelastic peak intensities. Increased
elastic intensity at $\vc{Q}$=(0~0~6) is due to the proximity to the
Bragg reflection.}\label{fig:polarization}
\end{figure}

Figure~\ref{fig:polarization}(a) shows the peak intensity change
with $\vc{Q}$ of the main-edge RIXS feature. The elastic intensity
shows a rapid drop above $\vc{Q}$=(0~0~3). The large elastic peak
intensity at $\vc{Q}$=(0~0~6) corresponds to the first Bragg
reflection. The inelastic intensities of M1 and M2 peaks start to
decrease after $\vc{Q}$=(0~0~6) and become quite small for higher
$\vc{Q}$. In contrast, Fig.~\ref{fig:polarization}(b) shows that the
intensities of the P1 and P2 peaks are more or less stable over the
whole $\vc{Q}$ range.

%
%

\begin{figure}[t]
\setlength{\unitlength}{\textwidth}
\begin{picture}(1,0.59)(0,0)
\put(0.01,-0.037){\includegraphics[width=0.45\textwidth,angle=0]{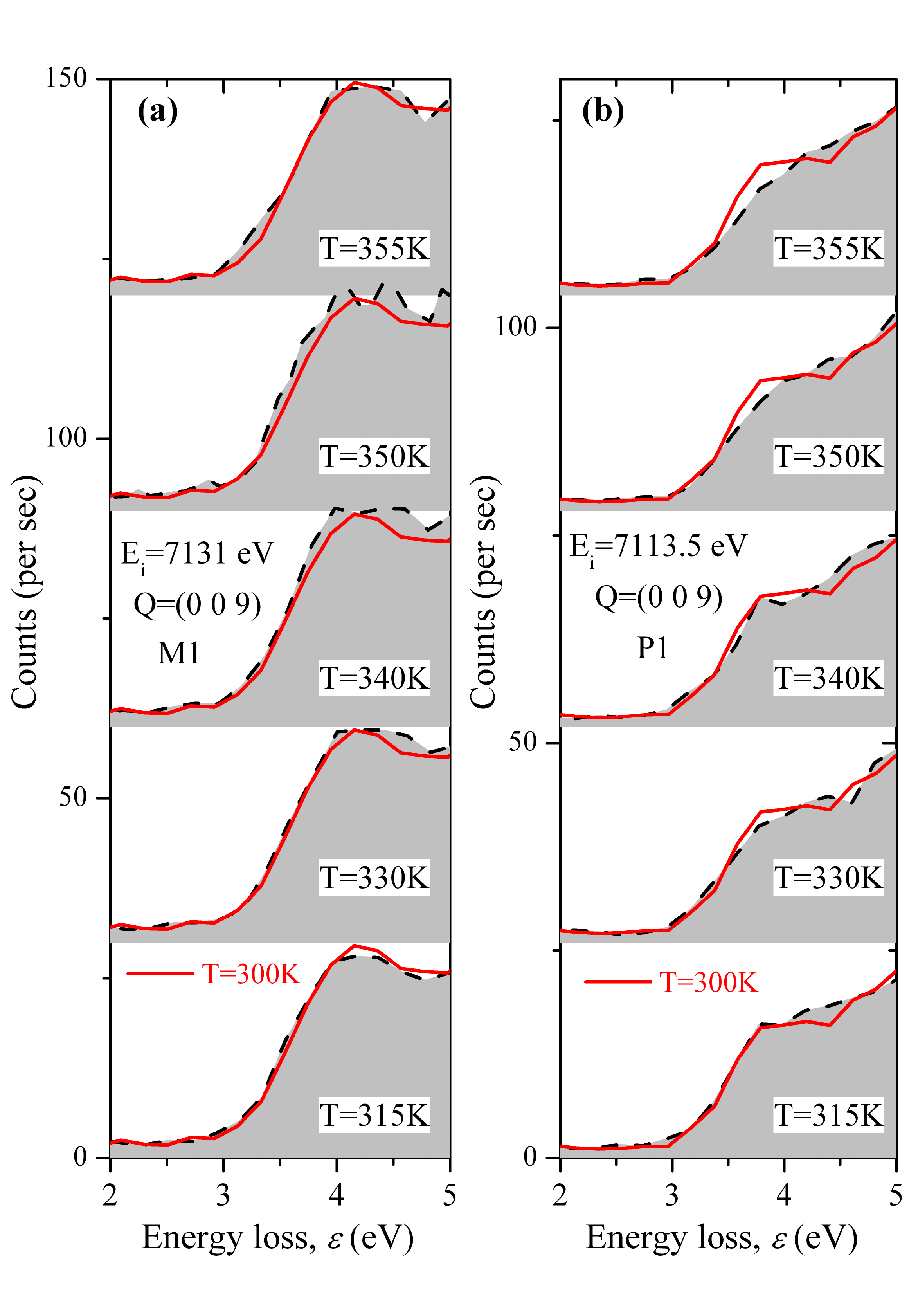}}
\end{picture}
\caption{(a) Main-edge RIXS ($E_{i}$=7131~eV, \vc{Q}=(0~0~9)) and
(b) pre-edge RIXS ($E_{i}$=7113.5~eV, \vc{Q}=(0~0~9)) spectra
measured at different crystal temperatures in the vicinity of the
N\'eel transition temperature $T_{\mathrm N}$=348~K.}
\label{fig:temp}
\end{figure}
\subsection{Temperature dependence}

To study the temperature dependence, the main-edge and pre-edge RIXS
spectra were measured at different crystal temperatures in the
vicinity of the N\'eel point, $T_{\mathrm N}$=348~K.
Figures~\ref{fig:temp}(a) and (b) show the main-edge and pre-edge
RIXS spectra, respectively, on the scale enlarged around the P1
peak. The measurements were performed at the momentum transfer
$\vc{Q}$=(0~0~9). For a better comparison, the RIXS spectra measured
at $T=300~K$ are superimposed on the spectra measured at higher
temperature. No significant systematic change is observed for the
main-edge RIXS spectra with temperature. The spectrum measured above
$T_{\mathrm N}$ at temperature $T=355~K$ matches well to the one
measured at $T=300~K$. On the other hand, the pre-edge RIXS spectrum
shows a noticeable change when crossing $T_{\mathrm N}$. Above
$T_{\mathrm N}$, the P1 peak becomes broad, resulting in a decreased
spectral intensity.

In optical spectroscopy, evidence has accumulated for the
sensitivity of MH excitations to local magnetic correlations, that
is, to the alignments of the nearest neighbor (NN) spins. For
example, LaMnO$_3$ is an $A$-type antiferromagnetic insulator, with
a ferromagnetic coupling along the $\mathbf{b}$-axis and an
antiferromagnetic coupling along the $\mathbf{c}$-axis. Upon
magnetic ordering, the spectral intensity of the lowest energy
excitation around 2~eV, which is associated with the MH excitation,
increases for the $\mathbf{b}$-axis but decreases for the
$\mathbf{c}$-axis.~\cite{Kovaleva04} Similar sensitivity of MH
excitations to local magnetic correlations was reported in a recent
RIXS study on a series of doped manganite systems.~\cite{Grenier05}
It was shown that the low energy RIXS spectral intensity increases
for FM nearest neighbor (NN) but decreases for paramagnetic (PM) NN.
On the other hand, the spectral intensity of the CT excitations is
known to be insensitive to local magnetic correlations, because the
oxygen 2$p$ states are completely filled. An often observed gradual
temperature dependence of CT excitation spectral intensity is
attributed to changes in the bond length or the orbital occupation.
No RIXS studies have evidenced the sensitivity of the CT excitations
to magnetic order. The different temperature dependences for MH and
CT excitations have been exploited to distinguish between these two
excitations in the optical spectroscopy.~\cite{Gossling08}

Using this analogy, one can suggest that no change in the main-edge
RIXS spectra is another argument that charge excitations probed by
this scattering process are CT excitations. The observed temperature
dependence in the pre-edge RIXS spectra suggests that charge
excitations probed by this scattering process are MH excitations.
The temperature dependence can be understood in the strong
crystal-field limit. As mentioned before, five $d$ electrons in \fb\
fill three-fold degenerate $t_{2g}$ and two-fold degenerate $e_g$ in
the high-spin configuration.  The Mott-Hubbard excitation in \fb\ is
the intersite charge transfer between two neighboring Fe ion sites
resulting in $d^4$ on one site and $d^6$ on the other site. In the
antiferromagnetic NN alignment, this transition occurs easily
because the NN site has the unoccupied level with the same spin with
the electron at the center site. On the other hand, in the
ferromagnetic NN alignment, the electron at the center site cannot
hop to the NN site because of the opposite spin state of the
unoccupied level at the NN site. Therefore, the change in the
antiferromagnetic NN alignment will be reflected in the MH
excitation.

\section{Discussion and Summary}

In the previous section, we presented the Fe K edge RIXS spectra in
\fb\ measured under different conditions, as a function of incident
photon energy, momentum transfer, photon polarization, and
temperature. The pre-edge and main-edge RIXS reveal different types
of charge excitations. The charge excitations probed by the
main-edge RIXS show a typical $\ei$ dependence observed in a number
of previous RIXS studies in
cuprates.~\cite{Abbamonte99,Lu06,yjkim07} Two low-lying M1 and M2
peaks show the dispersion widths of 0.26~eV and 0.44~eV,
respectively, and their intensities drop rapidly as $|\vc{Q}|$
increases. These excitations turned out to be insensitive to the
crystal temperature. On the contrary, the charge excitation revealed
by the pre-edge RIXS does not show any dispersion and any intensity
change with $\vc{Q}$. As the crystal temperature exceeds $T_{\mathrm
N}$, the P1 excitation with lowest energy broadens and decreases in
intensity. The observed temperature dependence is directly related
to local magnetic correlations. We attribute the main-edge and
pre-edge RIXS excitations to CT and MH excitations, respectively. In
the following, we discuss CT and MH excitations using molecular
orbitals (MO) in the cluster model and multiplet calculation in the
many-electron multiband model, respectively.

\begin{table}[b]
\caption {Peaks in the measured main-edge and pre-edge RIXS spectra
and their assignments in the framework of MO in the cluster model,
and multiplet calculation in the many-electron multiband model as in
Ref.~\onlinecite{OZ04}, respectively. The peak energies are measured
at $\vc{Q}$=(0~0~6). Energies in parenthesis are literature values
(Ref.~\onlinecite{Liechtenshtein82}).} \label{table1}
\begin{ruledtabular}
\begin{tabular}{ccl}
Peak & Assigned transition & Energy (eV)  \\
\hline\hline
\rm   Main-edge RIXS  & \\
\rm M1~~     & $t_{2u}(\pi)\rightarrow t_{2g}$    & 3.81\\
\rm M2~~     & $t_{1u}(\pi)\rightarrow t_{2g}$   & 4.94 (5.01) \\
\rm M3~\footnote{See the text for another plausible assignment.} & $t_{1u}(\sigma)\rightarrow e_g$   &8.14 (7.41)\\
\hline\hline
\rm Pre-edge RIXS   & $d^5:{^5}A_{1g}+d^5:{^5}A_{1g}~{\rightarrow}$\\
\rm P1~~ & $d^6:{^5}T_{2g}+d^4:{^5}E_{g}$    & 3.96 \\
\rm P2~~ & $d^6:{^5}E_{g}+d^4:{^5}E_{g}$     & 5.85 (5.36)  \\
\rm P3~~ & $d^6:{^1}A_{g}+d^4:{^1}T_{2g}$    & 8.71 (8.62)  \\
\end{tabular}
\end{ruledtabular}
\end{table}

\subsection{Charge-transfer excitation}

The ground state of undistorted octahedral Fe$^{3+}$O$_6$ of \fb\ is
described by filled non-bonding anionic ($O$~2$p$) MO
($a_{1g}(\sigma), t_{1g}(\pi), t_{1u}(\sigma), t_{1u}(\pi),$ and
$t_{2u}(\pi)$), filled bonding cationic MO ($t_{2g}(\pi)$ and
$e_g(\sigma)$), and unfilled anti-bonding 3$d$-type MO ($t_{2g}$ and
$e_g$).~\cite{Pisarev09} The CT excitation arises at the transition
from the anionic odd-parity MO ($t_{1u}(\sigma), t_{1u}(\pi),$ and
$t_{2u}(\pi)$) into the even-parity 3$d$-type MO ($t_{2g}$ and
$e_g$). Six CT excitations are allowed in the frame work of this
model. Among them, there are three strong dipole CT transitions
between MO states with the same bonding symmetry:
$t_{2u}(\pi)\rightarrow t_{2g}$, $t_{1u}(\pi)\rightarrow t_{2g}$,
and $t_{1u}(\sigma)\rightarrow e_g$.

Let us analyse the structure of the $t_{1u}(\sigma), t_{1u}(\pi),$
and $t_{2u}(\pi)$ energy states. The energy $E(t_{1u}(\sigma))$ of
$t_{1u}(\sigma)$ is lower than $E(t_{1u}(\pi))$ and $E(t_{2u}(\pi))$
due to its larger Madelung potential. Optical study suggests that
$E(t_{2u}(\pi))$ in \fb\ is higher than
$E(t_{1u}(\pi))$.~\cite{Markovin07} It is also known that
$E(t_{2u}(\pi))$ and $E(t_{1u}(\pi))$ are 2.2~eV and 1.2~eV higher,
respectively, than
$E(t_{1u}(\sigma))$.~\cite{Liechtenshtein82,Pisarev09}

In the MO model, therefore, the lowest CT excitation is the
transition from the highest occupied MO to the lowest unoccupied MO,
that is, $t_{2u}(\pi)\rightarrow t_{2g}$. The next one is
$t_{1u}(\pi)\rightarrow t_{2g}$. If we use the literature values
mentioned above, the energy difference between these two CT
excitations is 1.2~eV. This value is in agreement with the observed
mean energy difference between M1 and M2, which is about 1.13~eV.
Therefore, it is plausible to assign M1 and M2 excitations revealed
in the main-edge RIXS to $t_{2u}(\pi)\rightarrow t_{2g}$ and
$t_{1u}(\pi)\rightarrow t_{2g}$, respectively.

The highest CT excitation in this model is
$t_{1u}(\sigma)\rightarrow e_g$. Assuming that
$E(t_{2u}(\pi))-E(t_{1u}(\sigma))=$2.2~eV and
$E(e_{g})-E(t_{2g})\equiv10Dq=$1.4~eV, we can estimate that its
energy is 3.6~eV higher than the energy of the lowest
$t_{2u}(\pi)\rightarrow t_{2g}$ transition. This energy difference
of 3.6~eV has to be comparable with the observed energy difference
between the M1 and M3 excitations (4.31~eV). The discrepancy can be
attributed to some effects, such as an octahedral distortion, which
are not included in our cluster model. Therefore, we can assign the
M3 peak to the $t_{1u}(\sigma)\rightarrow e_{g}$ transition. Peak
assignments are summarized in Table~I.

The other plausible interpretation of the M3 excitation is due to MO
excitation from bonding to anti-bonding MO states. Taking into
account that bonding states are lower in energy than non bonding
states, this MO excitation can explain the larger energy separation
of M1 and M3. This type of MO excitation has been found in a number
of cuprates.~\cite{Yjkimprb04} It resonates at a higher incident
photon energy $E_i$, and its peak width is much larger than lower
energy CT excitations. Actually the $E_i$ dependence in
Figs.~\ref{fig:energydep}(c) and (d) shows that M3 resonates at
about 4~eV higher $E_i$ than M1 and M2. The peak width of M3 is much
broader compared with M1 and M2 (see Fig.~\ref{fig:energydep}(e)).

We have observed that M1 and M2 peaks show the dispersion widths of
0.26~eV and 0.44~eV, respectively. To fully understand this momentum
transfer dependence, more advanced calculations are needed beyond
the local cluster model. This would be very important for better
understanding the electronic structure of \fb.

\subsection{Mott-Hubbard excitation}

We will use the many-electron multiband model by Ovchinnikov and
Zabluda to analyze MH excitation in
FeBO$_3$.~\cite{Ovchinnikov03,OZ04} In this model, the Hamiltonian
of the system is written as
\begin{eqnarray}
H=\sum_{\lambda,\sigma}(\varepsilon_{\lambda}n_{\lambda
\sigma}+\frac{U_{\lambda}}{2}n_{\lambda \sigma}n_{\lambda
\overline{\sigma}})\;\;\;\;\;\;\;\;\;\;\;\;\;\;\; \nonumber
\\
+
\sum_{\frac{\lambda,\lambda'}{(\lambda\neq\lambda')}}\sum_{\sigma,\sigma'}(V_{\lambda
\lambda'}n_{\lambda \sigma}n_{\lambda' \sigma'}-J_{\lambda
\lambda'}a_{\lambda \sigma}^{\dagger}a_{\lambda \sigma'}a_{\lambda'
\sigma'}^{\dagger}a_{\lambda' \sigma}).
\end{eqnarray}
Here $\lambda$ and $\sigma$ are the orbital and the spin indices,
$a_{\lambda \sigma}$ ($a_{\lambda \sigma}^{\dagger}$) is the
creation (annihilation) operator of $d$ electrons with the spin
$\sigma$, and $n_{\lambda \sigma}=a_{\lambda
\sigma}^{\dagger}a_{\lambda \sigma}$. The quantity
$\varepsilon_{\lambda}$ is the atomic 3$d$ level energy that may
take the values of
$\varepsilon(t_{2g})=\varepsilon_d-0.4\times(10Dq)$ and
$\varepsilon(e_{g})=\varepsilon_d+0.6\times(10Dq)$. Coulomb
intraorbital repulsion energy $U_{\lambda}$ is nonzero when the 3$d$
electron occupies the same orbital with the different spin. Coulomb
interorbital repulsion energy $V_{\lambda \lambda'}$ is nonzero when
a 3$d$ electron occupies a different orbital. Hund exchange energy
$J_{\lambda \lambda'}$ is nonzero when a 3$d$ electron occupies a
different orbital with the same spin. Neglecting the orbital
dependence, these three energies are related by $U=2V+J$. Note that
this is an atomic model, ignoring 3$d$ spin-orbit interaction and
inter atomic exchange.

The ground-state crystalline term of \fb\ is $^{6}A_{1g}$. Its
energy $E(d^5$:$^{6}A_{1g})$ is equal to
$3\varepsilon(t_{2g})$+2$\varepsilon(e_{g})$+10$V$-10$J$ =
5$\varepsilon_d$+10$V$-10$J$. The lowest $d^4$ configuration
corresponds to occupied three $t_{2g}$ and one $e_g$ states in the
high spin configuration in which the crystalline term is
$^{5}E_{g}$. Its energy $E(d^4$:$^{5}E_{g})$ is equal to
$3\varepsilon(t_{2g})$+$\varepsilon(e_{g})$+6$V$-6$J$=4$\varepsilon_d$-$0.6\times$10$Dq$+6$V$-6$J$.
The lowest $d^6$ configuration corresponds to occupied four $t_{2g}$
and two $e_g$ in the high spin configuration in which the
crystalline term is $^{5}T_{2g}$. Its energy $E(d^6$:$^{5}T_{2g})$
is equal to
$4\varepsilon(t_{2g})$+2$\varepsilon(e_{g})$+$U$+14$V$-10$J$=6$\varepsilon_d$-$0.4\times$10$Dq$+$U$+14$V$-10$J$.
Now we can calculate the lowest MH excitation energy
$U_{\mathrm{G}}=E(d^6$:${^5}T_{2g})$+$E(d^4$:${^5}E_{g})-$2$E(d^5$:${^6}A_{1g})$=$U+4J-10Dq$.

Higher MH excitations arise from excited $d^4$ and $d^6$
configurations. Crystalline terms of excited $d^4$ configurations
are $^3T_{1g}$ and $^1T_{2g}$ in the order of increasing energy,
while those of excited $d^6$ configurations are $^5E_{g}$,
$^1A_{1g}$, $^3T_{1g}$ and $^3T_{2g}$. Considering the
spin-selection rule, higher MH excitations are
$E(d^6:{^5}E_{g})+E(d^4:{^5}E_{g})-2E(d^5:{^6}A_{1g})
=U_{\mathrm{G}}+10Dq$ and
$E(d^6:{^1}A_{1g})+E(d^4:{^1}T_{2g})-2E(d^5:{^6}A_{1g})
=U_{\mathrm{G}}+2U-2V+6J-2\times10Dq$ in the order of increasing
energy.

We assign the P1 peak in the pre-edge RIXS spectra to the lowest MH
excitation and obtain $U_{\mathrm{G}}=U+4J-10Dq$=3.96~eV. The
literature values for $J$ are spread from 0.35~eV to
1~eV.~\cite{Ovchinnikov03,OZ04,Shang07} Based on optical studies,
Ovchinnikov and Zabluda~\cite{OZ04} obtained $U=2.97$~eV using
$J=$0.7~eV and $10Dq=1.57$~eV. In our case, we are measuring
$10Dq=1.4$~eV (see Fig.~\ref{fig:energydep}(b)) and obtain
$U=2.56$~eV assuming $J=$0.7~eV.

The second MH excitation energy is equal to 5.36~eV, which is
10$Dq$=1.4~eV higher than the lowest one. We observe that the mean
energy difference between P1 and P2 is about 1.85~eV. We assign the
P2 peak to the second MH excitation:
$(d^5:{^6}A_{1g})+(d^5:{^6}A_{1g}){\rightarrow}(d^6:{^5}E_{g})+(d^4:{^5}E_{g})$.
The somewhat large discrepancy in energy can be attributed to
various effects, such as 3$d$ spin-orbit interaction and inter
atomic exchange, which are not included in the model. In addition,
the energy estimation is based on the local multiplet model where
the inter site excitation state is a simple product of two single
site states and inter site mixing is not included. In this regard,
more refined model calculations are desirable. In the case of the
third MH excitation, the calculated energy equals to 8.62~eV. Since
the energy of the P3 excitation is 8.71~eV, we can assign the P3
excitation to the third MH excitation:
$(d^5:{^6}A_{1g})+(d^5:{^6}A_{1g}){\rightarrow}(d^6:{^1}A_{1g})+(d^4:{^1}T_{2g})$.
Peak assignments are summarized in Table~I.

\subsection{Electronic structure in \fb}

We have assigned all observed peaks in the measured RIXS spectra
using molecular orbitals (MO) in the cluster model and multiplet
calculation in the many-electron multiband model. Using these
assignments, we can derive the electronic structure around the
chemical potential ($\mu$) in \fb. The lowest UHB energy is
$E(d^6$:${^5}T_{2g})-E(d^5$:${^6}A_{1g})=\varepsilon_d-0.4\times{10Dq}+U+V=\varepsilon_d+5.72$~eV
and the highest LHB energy is
$E(d^4$:${^5}E_{g})-E(d^5$:${^6}A_{1g})=-\varepsilon_d-0.6\times{10Dq}-4V+4V=-\varepsilon_d-1.76$~eV,
resulting in $\gapmh$=3.96~eV. On the other hand, when
$-\varepsilon_p$ is introduced to describe the non bonding (NB) $O$
2$p$ state relative to $\mu$, the lowest CT energy is
$\varepsilon_d+5.72-\varepsilon_p$~eV, resulting in
$\gapct$=3.81~eV.

\begin{figure}[t]
\setlength{\unitlength}{\textwidth}
\begin{picture}(1,0.3)(0,0)
\put(0.01,-0.05){\includegraphics[width=0.42\textwidth,angle=0]{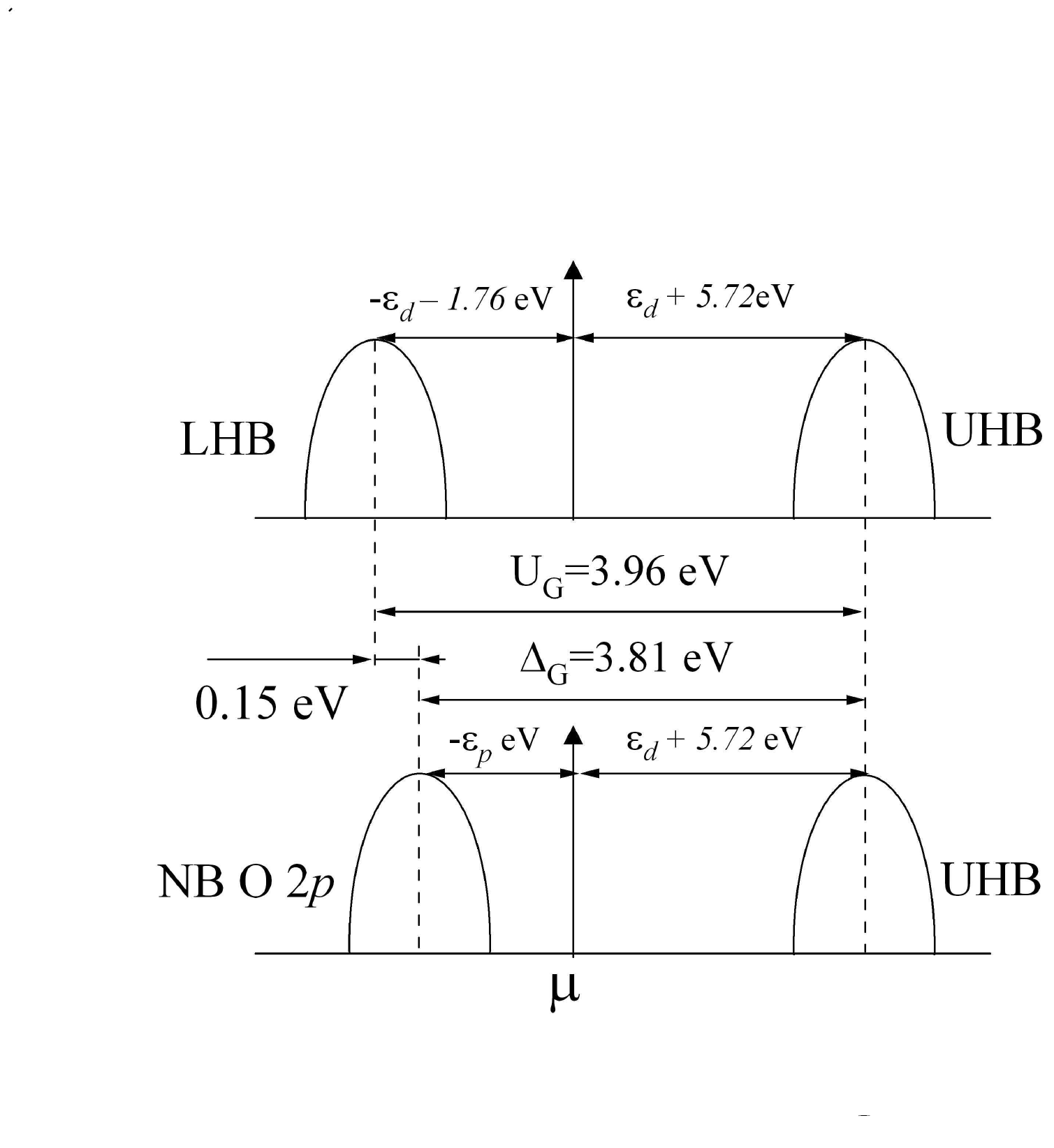}}
\end{picture}
\caption{Schematic of electronic structure around the chemical
potential ($\mu$) in \fb. Energies are in the unit of eV, based on
the results of the presented RIXS measurements.} \label{fig:band}
\end{figure}

The schematic in Fig.~\ref{fig:band} presents the electronic
structure in \fb\ around $\mu$ based on our RIXS studies. Our
experimental results support the conclusion that \fb\ is a CT
insulator, as found by Ovchinnikov and Zabluda~\cite{OZ04} based on
optical absorption and photoemission spectroscopy studies. However,
the values for the energy separation between NB O~2$p$ and LHB are
quite different: 0.15~eV in the present study, and 1.4~eV in the
study reported by Ovchinnikov and Zabluda. We believe that our
result is more accurate because it was obtained, model independent
using the same spectroscopy to probe both CT and MH excitations.

%
%

\section{Summary}

In this study, we report the first successful resonant inelastic
x-ray scattering (RIXS) measurement at the K absorption edge of Fe.
We observe qualitatively different RIXS excitations associated with
the main-edge and pre-edge absorption phenomena. These two types of
excitations have different incident energy, polarization, and
momentum dependence. We interpret the main-edge and pre-edge RIXS
excitations as charge-transfer and Mott-Hubbard excitations,
respectively. We assign and discuss CT and MH excitations using
molecular orbitals (MO) in the cluster model and multiplet
calculation in the many-electron multiband model, respectively. The
present study demonstrates the utility of $K$ edge RIXS as a
simultaneous probe of charge-transfer and Mott-Hubbard excitations
in transition metal compounds. Novel phenomena of the
insulator-semiconductor transition and the collapse of the magnetic
moment have been observed recently in \fb\ under
pressure.~\cite{Troyan03,Sarkisyan02,GTO04} For a detailed
understanding of these phenomena, it is essential to know
charge-transfer energy, crystal-field splitting $10Dq$, and Coulomb
repulsion $U$. We have demonstrated that K edge RIXS can be used to
measure these quantities directly. K edge RIXS, being a hard x-ray
bulk-sensitive probe, can be applied for high-pressure studies. It
would be very appealing in the next step to use these techniques to
study the pressure-induced electronic and magnetic property changes
in \fb.

\acknowledgements{Yu. Sh. acknowledges the longstanding effort of
his colleagues from the IXS Collaborative Design Team in building
MERIX instrument at the XOR-IXS 30-ID beamline at the APS, in
particular of John Hill, Scott Coburn (BNL), Clement Burns (WMU),
Ercan Alp, Thomas Toellner, and Harald Sinn (APS). He is also
indebted to Ruben Khachatryan (APS), Michael Wieczorek (APS), and
Ayman Said (APS) for the help in manufacturing the Ge(620) analyzer.
Peter Siddons (BNL) is acknowledged for building the microstrip
detector for the MERIX spectrometer. The help of the XOR-IXS 30-ID
beamline personnel at the Advanced Photon Source: Tim Roberts, Ayman
Said, Mary Upton is greatly appreciated. Use of the Advanced Photon
Source was supported by the U. S. DOE, Office of Science, Office of
Basic Energy Sciences, under Contract No. DE-AC02-06CH11357.}

\bibliography{febo3rixs}

\end{document}